\documentclass{PoS}

\newcommand{\trc}{{\rm Tr_c~}}

\newcommand{\re}{{\rm Re~}}

\def\lsim{\raise0.3ex\hbox{$<$\kern-0.75em\raise-1.1ex\hbox{$\sim$}}}
\def\gsim{\raise0.3ex\hbox{$>$\kern-0.75em\raise-1.1ex\hbox{$\sim$}}}
\def\bx{{\bf x}}

\newcommand{\non}{\nonumber}
\newcommand{\be}{\begin{enumerate}}
\newcommand{\ee}{\end{enumerate}}
\newcommand{\bi}{\begin{itemize}}
\newcommand{\ei}{\end{itemze}}
\newcommand{\beq}{\begin{equation}}
\newcommand{\eeq}{\end{equation}}
\newcommand{\beqn}{\begin{eqnarray}}
\newcommand{\eeqn}{\end{eqnarray}}
\newcommand{\lla}{\left\langle}
\newcommand{\rla}{\right\rangle}

\title{Correlations of chiral condensates and quark number densities with static quark sources}

\ShortTitle{Correlations of chiral condensates and quark number densities with static quark sources}

\author{\speaker{Kay H\"ubner}\thanks{for the RBC-Bielefeld Collaboration, together with C.~Schroeder, UC San Diego}\\
        Brookhaven National Lab\\
        E-mail: \email{huebner@bnl.gov}}

\abstract{
We investigate correlation functions of the Polyakov loop and static meson/diquark systems with the chiral condensate and the quark number density at finite temperature. In particular the latter observable can give insight into the mechanism of screening and string breaking at finite temperature. We use for our analysis gauge field configurations generated in 2+1 flavor QCD with an improved staggered fermion action with almost physical light quark masses and a physical value of the strange quark mass on lattices with temporal extent $N_\tau=4$ and $6$.
}

\FullConference{The XXV International Symposium on Lattice Field Theory\\
		 July 30-4 August 2007\\
		 Regensburg, Germany}

\begin{document}

\section{Introduction}

Global expectation values of the chiral condensate and quark number density give information about the structure of the thermal medium at finite temperature and non-zero chemical potential $\mu$.
Calculating these observables {\em locally}, i.~e.~depending on their spatial position, 
allows for a spatially resolved analysis of dynamical quark clouds and the chiral condensate around static quark, diquark or Meson sources.
We investigate correlation functions of the chiral condensate/quark number density with the Polyakov loop or products of Polyakov loops, which  allow to probe the screening and string breaking mechanism of systems composed of static color charges at finite temperature.
Lattice calculations of these operators have been performed in the past with staggered and Wilson quarks \cite{Feilmair:1989hi,Feilmair:1988js,Sakuler:1992qx,Buerger:1994zc} for rather large quark masses and on small lattices.
The investigations presented here form the basis for future studies of the spatial correlators of the quark number density, which has been proposed as a means to decide whether the QGP has a liquid phase \cite{Thoma:2005aw}.

\section{Operators}


We have calculated the chiral condensate and the quark number density depending on their spatial position $\bf x$ by using 
%
%
\beq
   \bar\chi\chi^f ({\bf x})=\frac{n_f}{4N_\tau} \trc \sum_{x_0}{\bf M}^{-1}_{f,xx}\quad\mbox{and}\quad
  n^f_q ({\bf x})= \frac{n_f}{4N_\tau} \trc \sum_{y,x_0}{\bf M}^{-1}_{f,xy}\frac{\partial {\bf M}_{f,yx}}{\partial\mu_f},
\label{eq:nq}
\eeq
where $x=(x_0,\bx)$, $f=1,\dots,n_f$ labels the flavor, $\mu_f$ is the corresponding chemical potential and $\trc$ denotes the trace in color space, with $\trc{\mathbf 1}=1$. Derivatives with respect to  $\mu_f$ are evaluated at $\mu_f=0$. 
%
%
%
%
We can now calculate the correlator of $\bar\chi\chi^f ({\bf x})$ and $n^f_q ({\bf x})$ with the Polyakov loop $L({\bf x})$ or products of Polyakov loops, for which we use  
\beq
\non
   L({\bf x})=Z(g^2)^{N_\tau}\,\trc\prod_{x_0=0}^{N_\tau-1} U_4(x_0,\bx),\quad C_{Q\bar Q}(r)=L(0)L^\ast(r),\quad C_{QQ}(r)=L(0)L(r),
\eeq
where $L$, and therefore also $C_{Q\bar Q}$ and $C_{QQ}$, is properly renormalised  by the multiplicative renormalisation constant $Z(g^2)^{N_\tau}$, which is determined at $T=0$ from a matching of the static quark potential to the string potential. 
The $C_{Q\bar Q}$ and $C_{QQ}$  describe static meson and diquark systems, respectively.
The correlation functions $\lla n_q^f({\bf x})L({\bf y})\rla$ and $\lla n_q^f({\bf x})C_s(r)\rla$ for $s=QQ,Q\bar Q$ are then properly renormalised as well. 
The renormalisation for $\bar\chi\chi^f$ involves both multiplicative and additive renormalisation constants. In order to eliminate the additive contributions, we calculate the {\em connected} correlator
\beq
\label{eq:pbp_connected}
\lla \bar\chi\chi^f({\bf x})L({\bf y})\rla_{c}=\lla \bar\chi\chi^f({\bf x})L({\bf y})\rla-\lla \bar\chi\chi^f\rla \lla L\rla,
\eeq
which approaches the corresponding continuum expression  $\lla \bar\psi\psi^f({\bf x})L({\bf y})\rla_c$ in the limit $a\to 0$. 

In this work, we analyze gauge field configurations generated in 2+1 flavor QCD 
with the p4fat3 staggered fermion action, where the strange quark mass $m_s$ is fixed to its physical value and the light quark mass is choosen as $m_l=0.1m_s$. Lattice sizes are $16^3\times 4,6$ \cite{EoS,Cheng:2006qk}. For the calculation of expectation values we employed the random noise vector method with $100$ random noise vectors.

\section{Results for $\lla \bar\chi\chi L \rla_c$ and $\lla n_q L\rla$}

\begin{figure}[t]
\scalebox{0.6}{\includegraphics{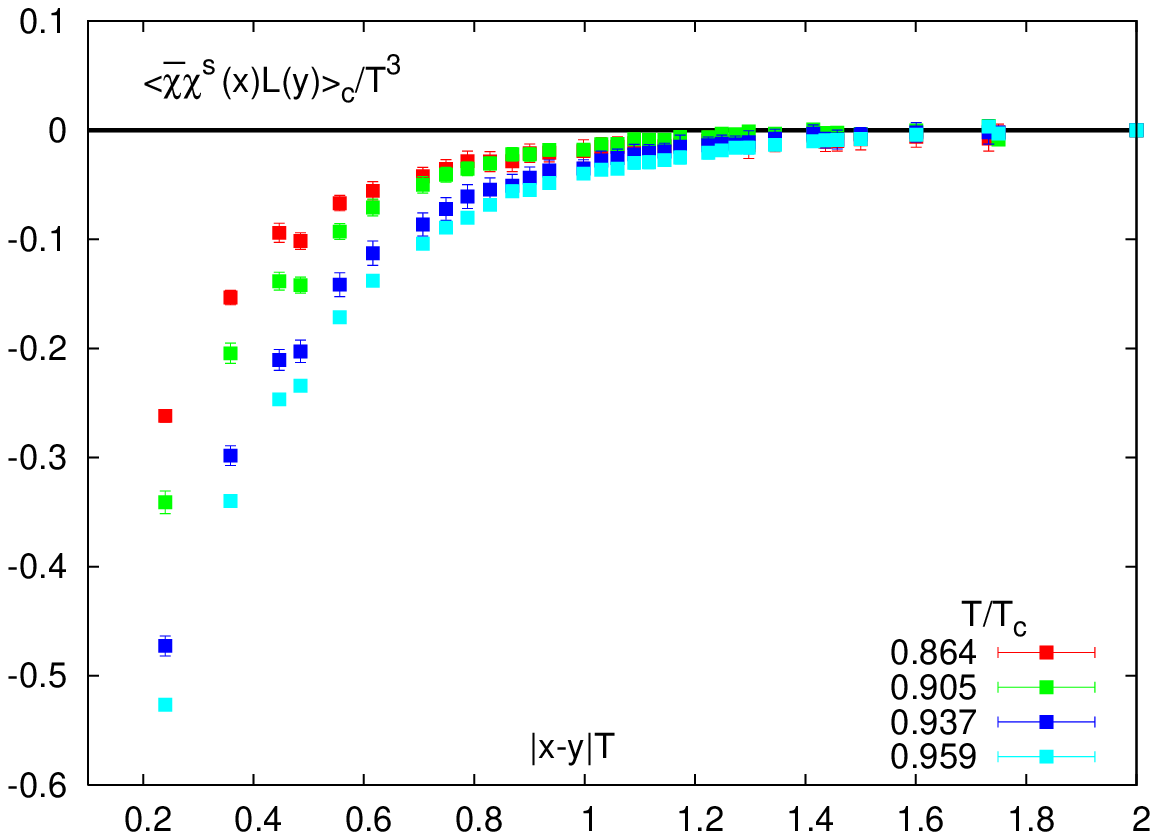}}
\scalebox{0.6}{\includegraphics{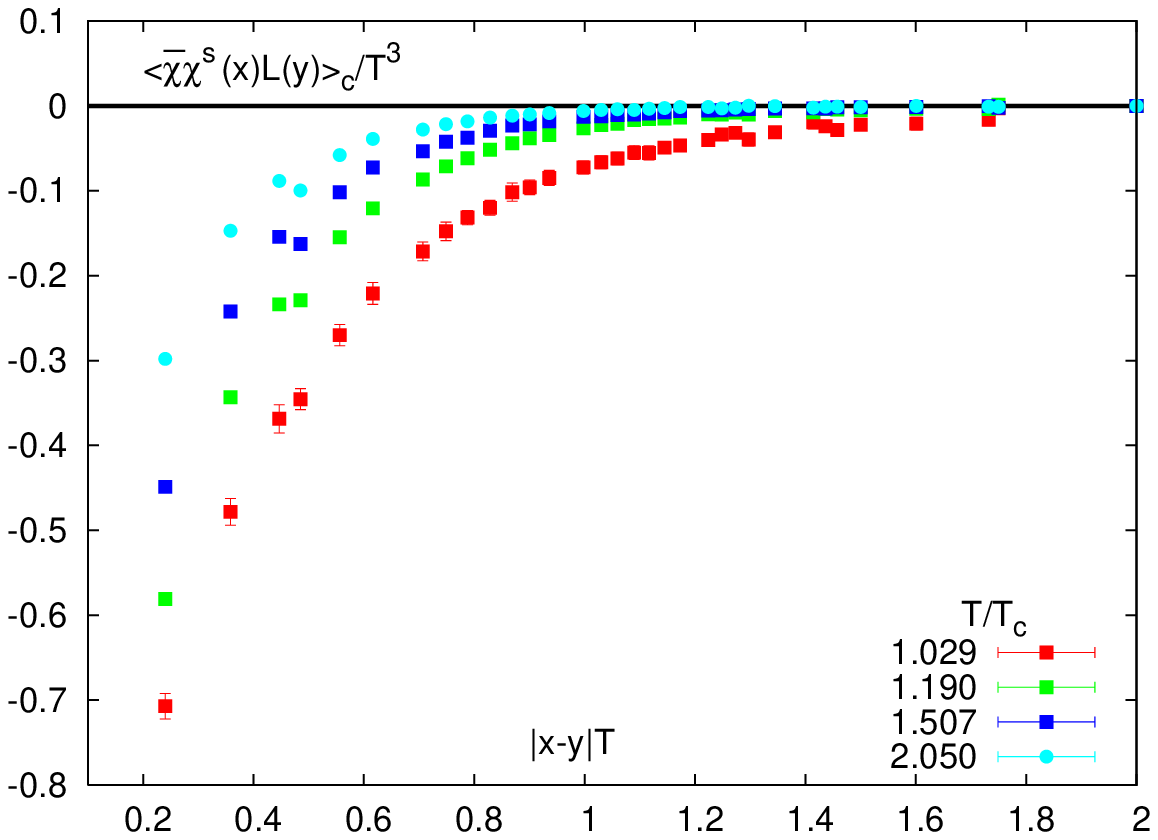}}
\scalebox{0.6}{\includegraphics{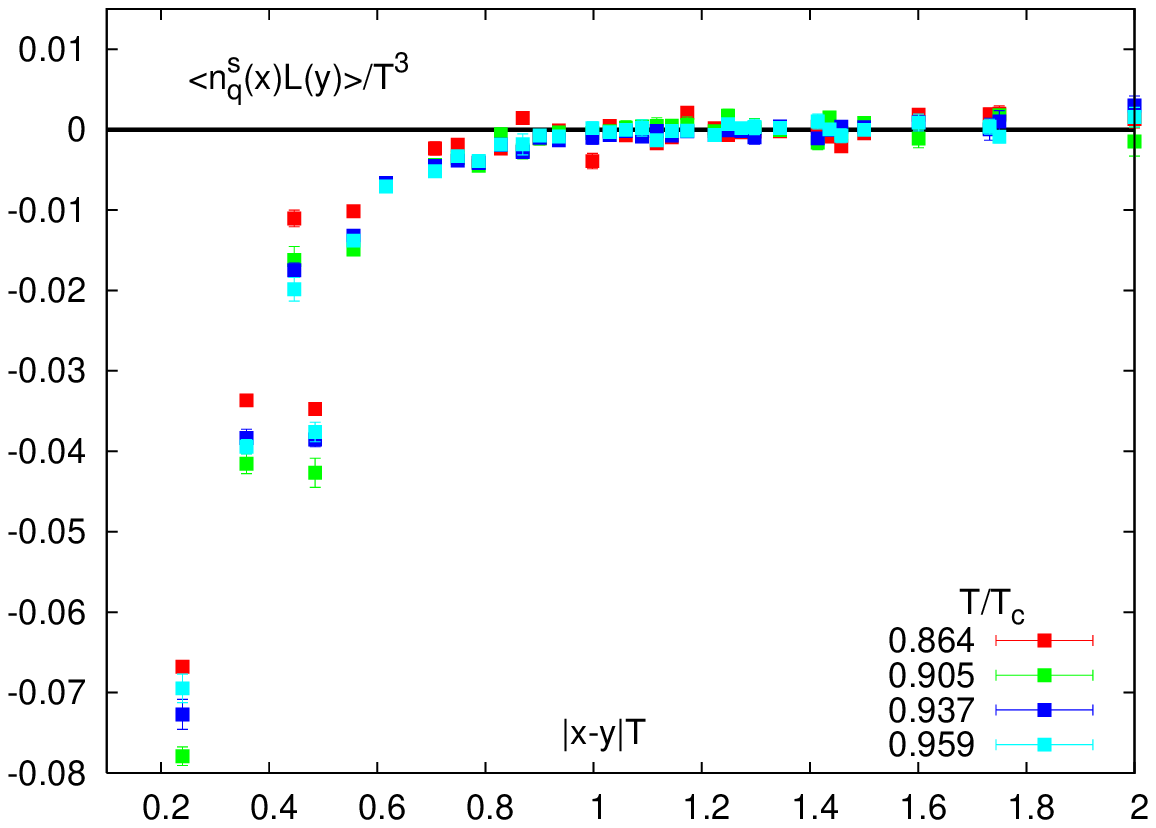}}
\scalebox{0.6}{\includegraphics{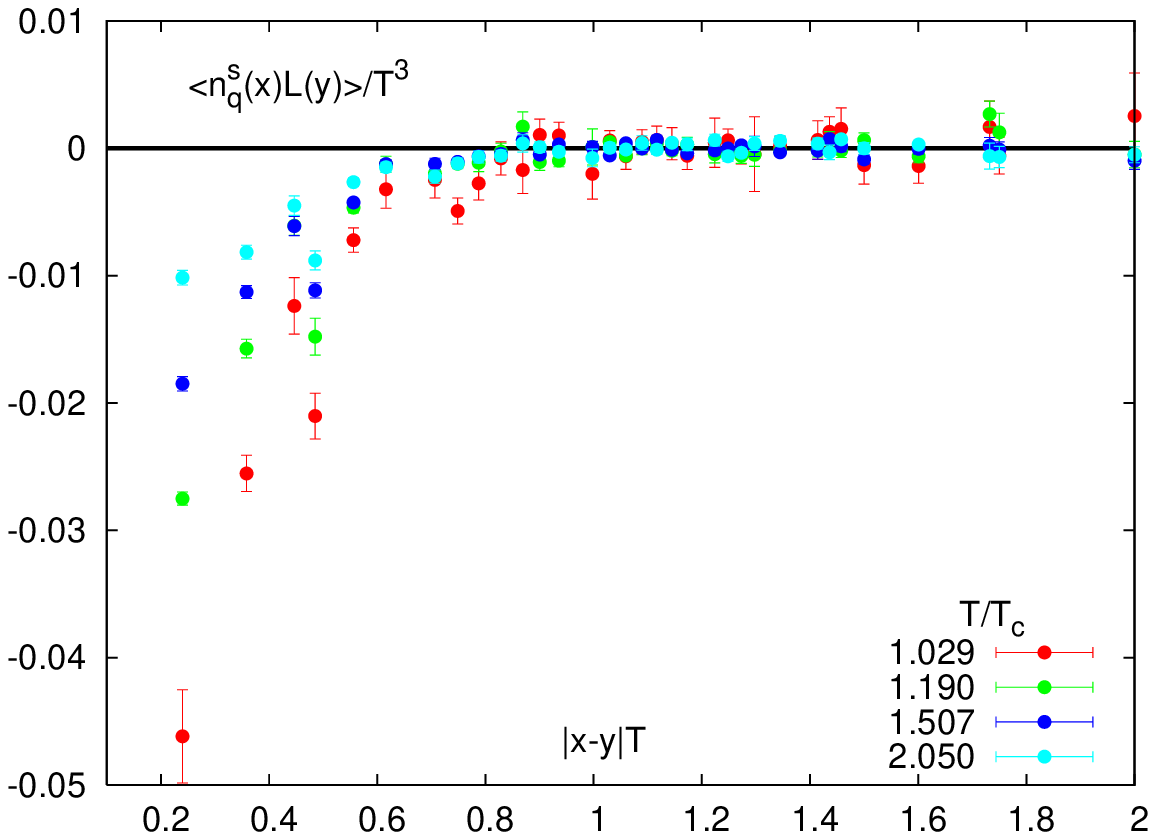}}
\caption{\label{fig:xL}$\lla \bar\chi\chi^s L \rla_c$ (upper row) and $\lla n_q^s L\rla$ (lower row) for the strange quark species.
}
\end{figure}
 In fig.~\ref{fig:xL} we show the results for $\lla \bar\chi\chi^s L \rla_c$ (upper row) and $\lla n_q^s L\rla$ (lower row) for the strange quark species obtained from $N_\tau = 4$ lattices.
We observe that all correlators vanish rapidly at distances $rT\gsim 1$,
i.~e.~when $\bar\chi\chi^f$ and $n_q^f$ approach their thermal expectation value 
undisturbed by the presence of the static quark.  For small separations $\lla \bar\chi\chi^s L \rla_c$  assumes smaller values, signalling a gradual restoration of chiral symmetry in the vincinity of the static quark. 
The correlator  $\lla n_q^s L\rla$ becomes smaller than zero for smaller separations as well, as the static quark source  is screened by an excess of dynamical anti-quarks arranged in a cloud around the static quark.  The effect becomes smaller above $T_c$, as thermal gluons are available in the deconfined phase to provide screening.  All these findings are in qualitative agreement with those of earlier studies \cite{Feilmair:1989hi,Feilmair:1988js,Sakuler:1992qx,Buerger:1994zc}.

\begin{figure}[t]
\scalebox{0.6}{\includegraphics{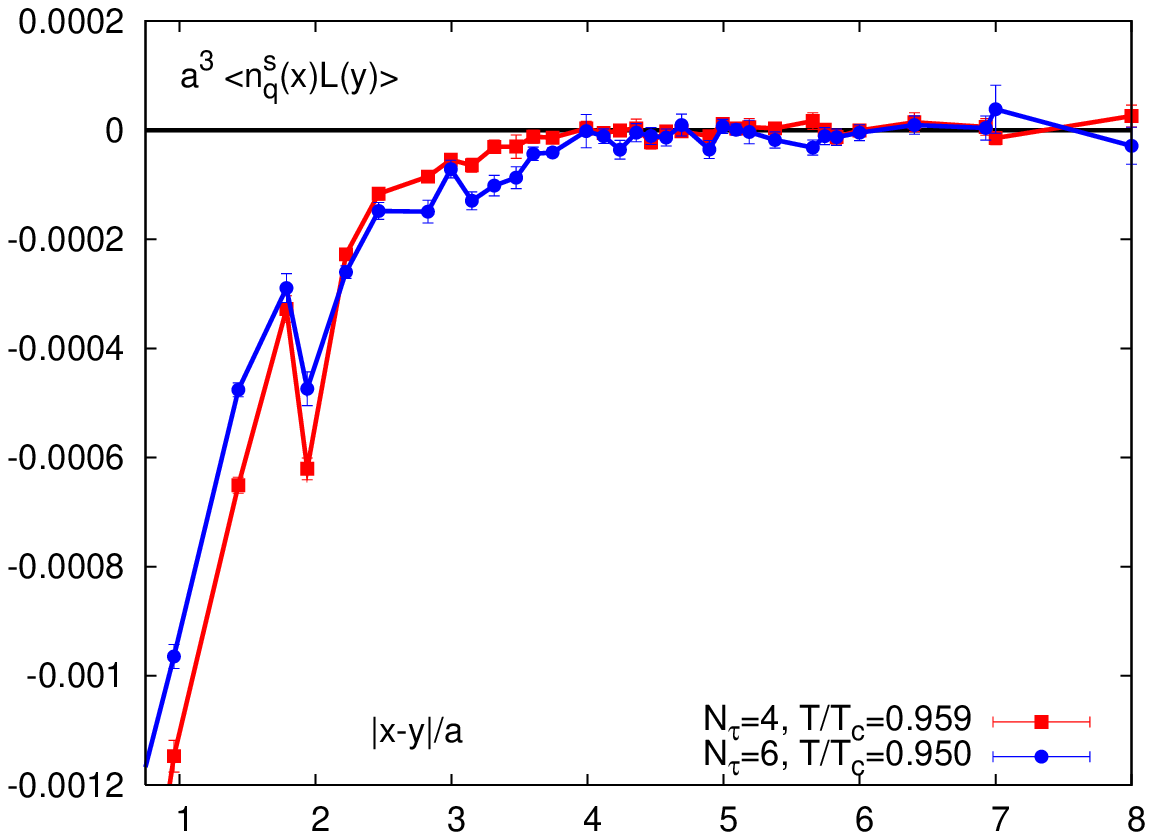}}
\scalebox{0.6}{\includegraphics{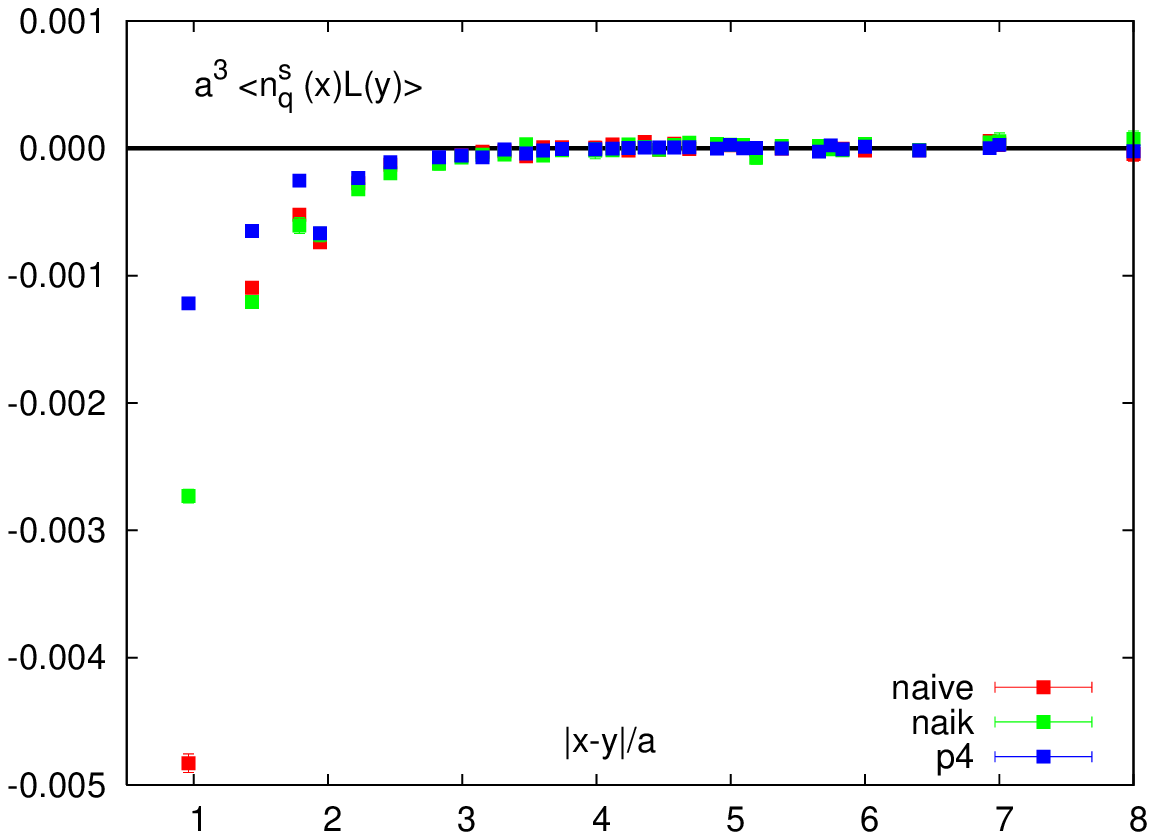}}
\caption{\label{fig:artifact}$\lla n_q^s L\rla$ for $N_\tau=4$ and 6 (left) and for different staggered discretisations of the fermion matrix $\bf M$ at $T/T_c=0.905, N_\tau=4$ (right).
}
\end{figure}

For $\lla n_q^s L\rla$ we observe a peculiar dip at $rT \approx 0.5$, which we investigate in more detail in fig.~\ref{fig:artifact}. Varying the lattice spacing, we see in fig.~\ref{fig:artifact} (left), that the position of the dip varies with $a$ as well, pointing at a lattice artifact that vanishes in the continuum limit.
In fig.~\ref{fig:artifact} (right)  we computed $\lla n_q^s L\rla$ at $T/T_c=0.905, N_\tau=4$ for p4, Naik and naive staggered discretisations of the fermion matrix $\bf M$. 
The artifact is present in all three discretisation schemes, thus its appearance is  not dependent on the particular action used.  
Rather the mixing of different quantum channels in the staggered formulation is probably responsible for this effect. 

In order to quantify further the behavior of these correlators, we have performed simple exponential fits.
%
%
%
\begin{figure}[h]
\scalebox{0.6}{\includegraphics{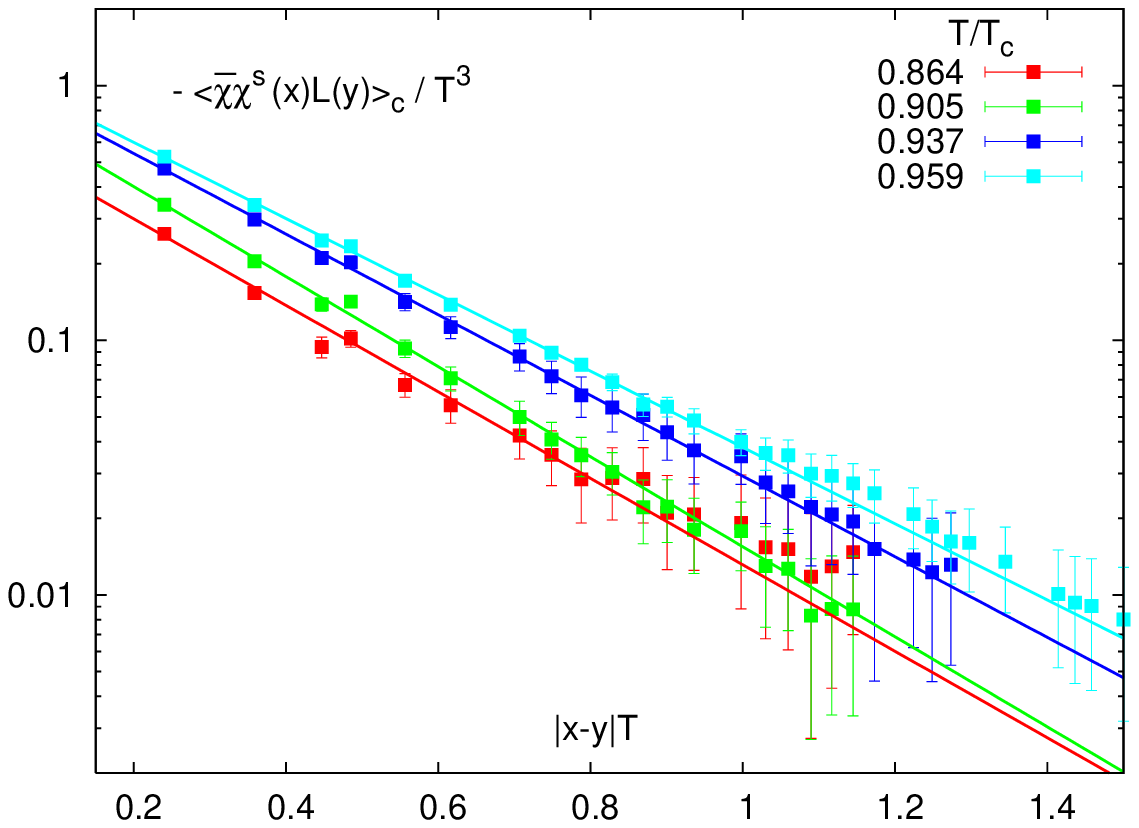}}
\scalebox{0.6}{\includegraphics{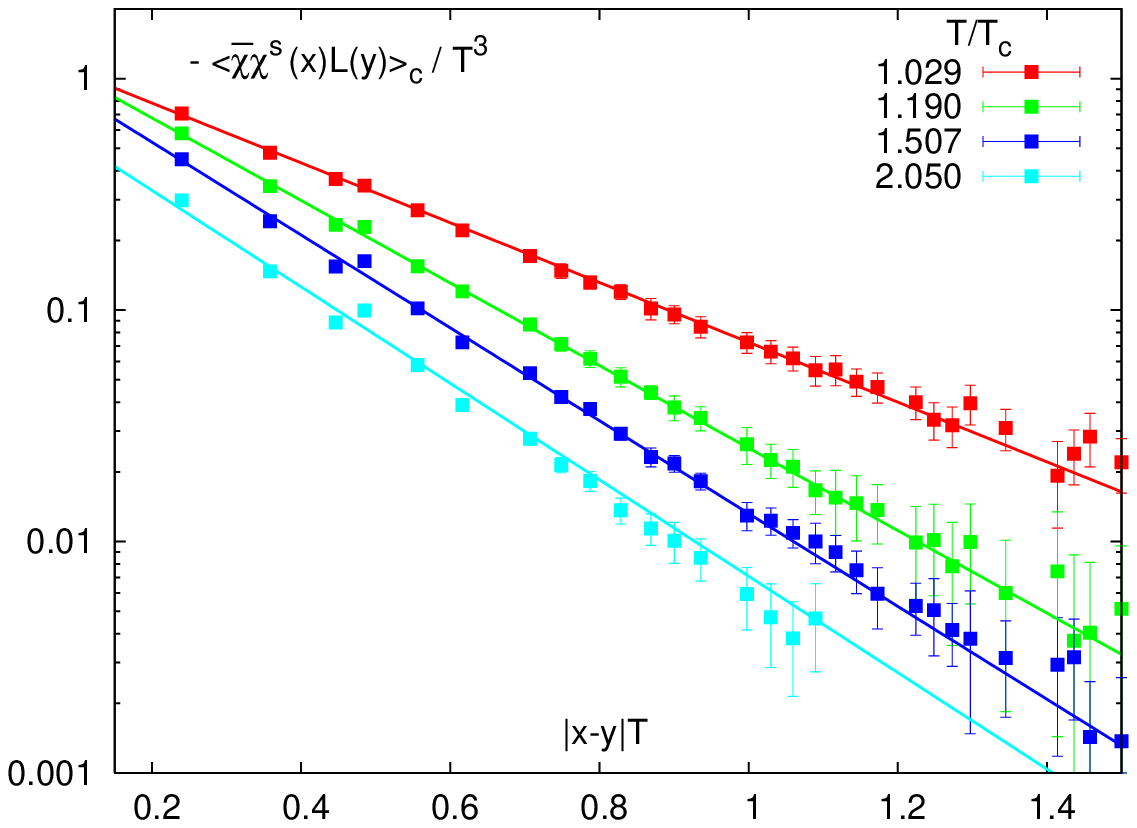}}
\scalebox{0.6}{\includegraphics{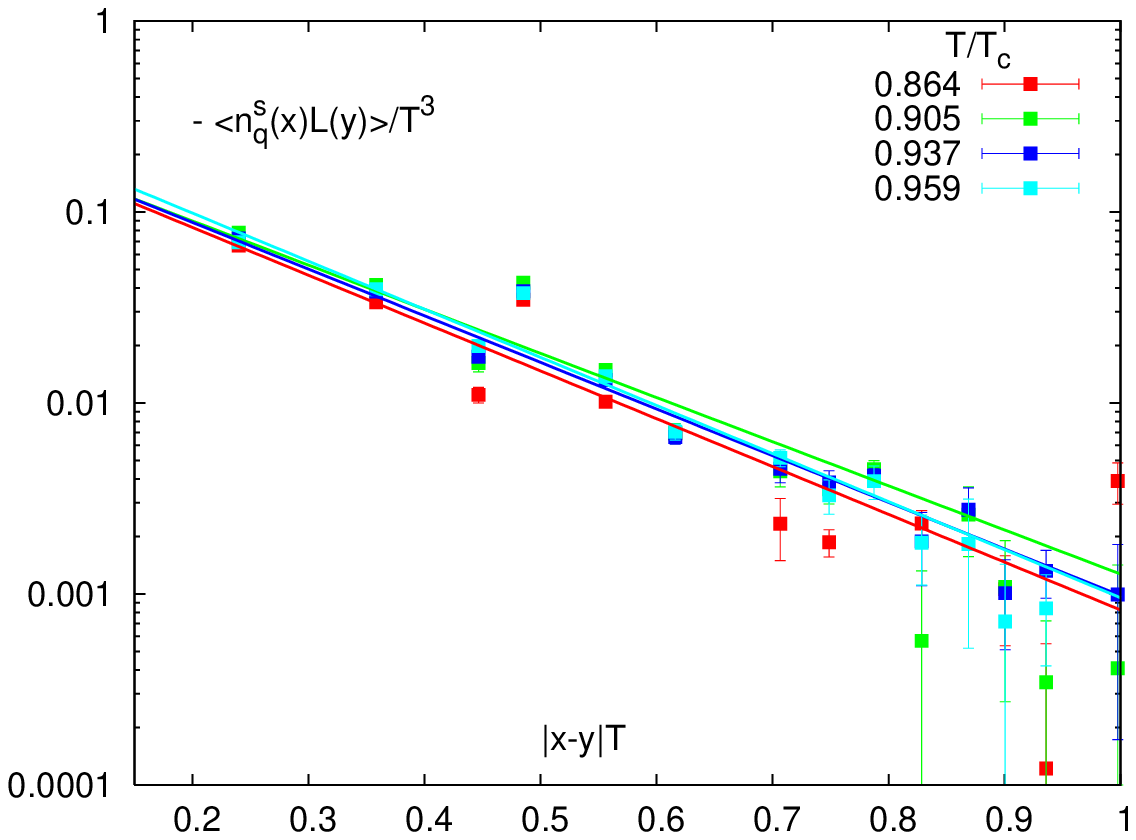}}
\scalebox{0.6}{\includegraphics{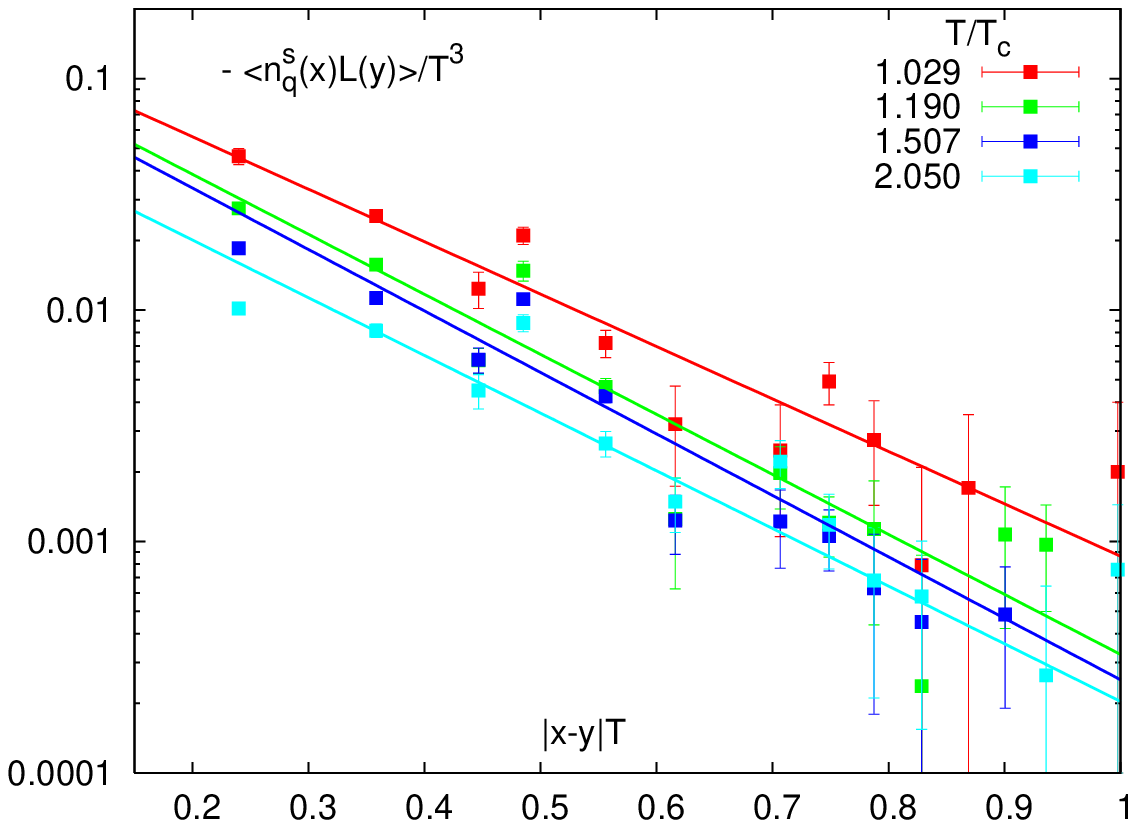}}
\caption{\label{fig:xL_log} $\lla \bar\chi\chi^s L \rla_c$ (upper row) and $\lla n_q^s L\rla$ (lower row) for the strange quark species on a log-scale.  Lines are the results of  the $\chi^2$-fits with simple exponential functions.
}
\end{figure}
\begin{figure}[h]
\scalebox{0.6}{\includegraphics{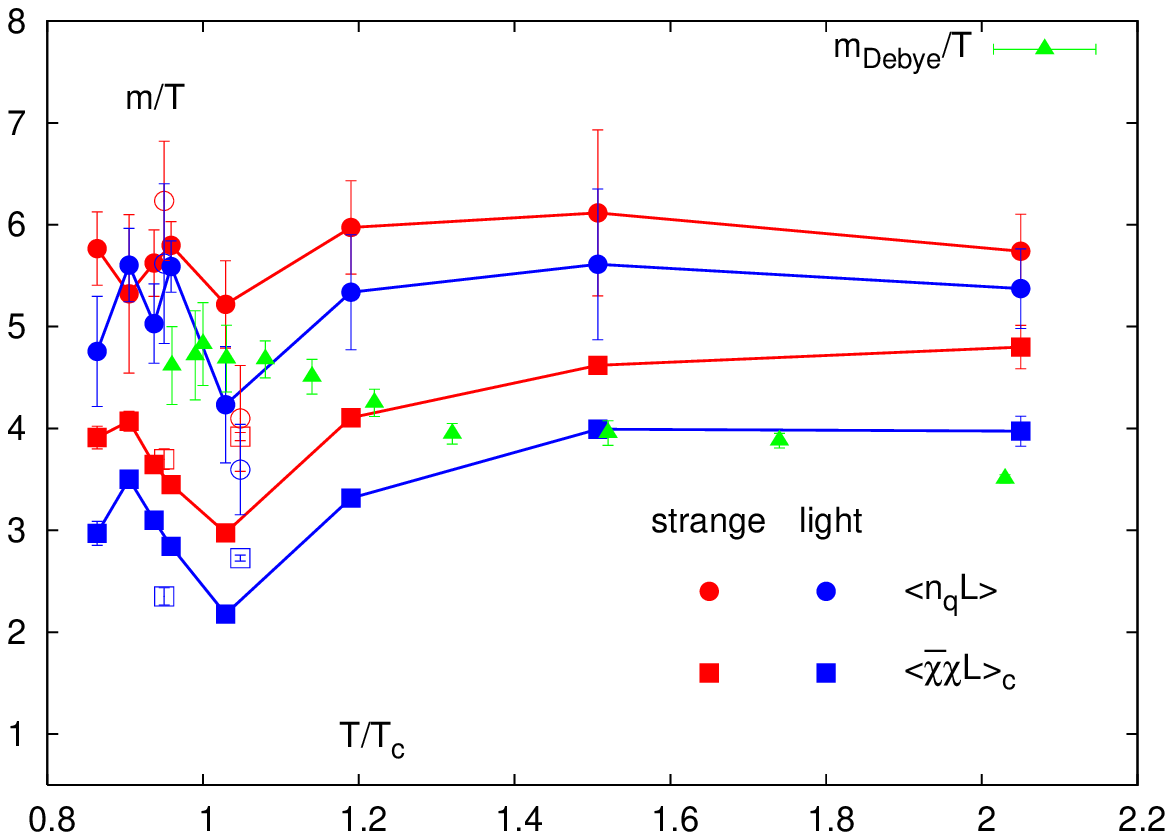}}
\scalebox{0.6}{\includegraphics{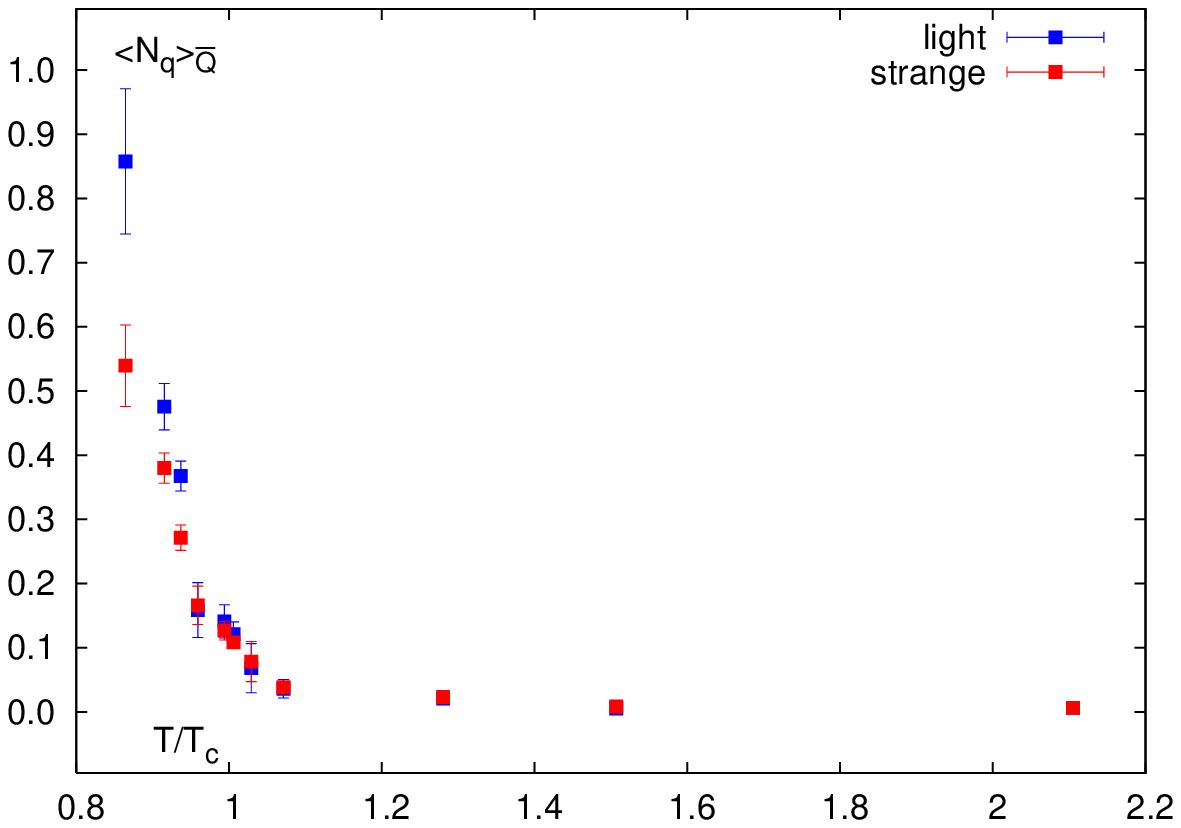}}
\caption{\label{fig:screening_masses}Left: Screening masses obtained from an exponential fit. 
Closed symbols: $N_\tau=4$, open symbols: $N_\tau=6$. Debye masses are taken from \cite{kostya}.
Right: Induced quark number around a static anti-quark.
}
\end{figure}
The resulting screening masses $m^{l,s}$ of the correlator involving the chiral condensate are, of course, not affected by the multiplicative renormalisation constants.
In fig.~\ref{fig:xL_log} we show $\lla \bar\chi\chi^s L \rla_c$ (upper row) and $\lla n_q^s L\rla$ (lower row) for the strange quark species on a log-scale together with the results of  the $\chi^2$-fits of exponential functions. 
We obtain similar results for the light quark species.
The fit ranges employed were $0.3\le rT\le1$ for $\lla n_q^f L\rla$, where we left out the values due to the artifact at $rT\approx 0.5$, and $0.1\le rT\le1.2$ for $\lla \bar\chi\chi^f L \rla_c$.

The results for the screening masses obtained from the fits with exponential functions are shown in fig.~\ref{fig:screening_masses}(left). 
As expected\footnote{For heavy sea quarks the correlation functions $\lla \bar\chi\chi^f L \rla_c$ and $\lla n_q^f L\rla$ can be related to hopping parameter expansions of real and imaginary parts of Polyakov loops, respectively\cite{Feilmair:1988js}. In this limit one finds at high temperatures $m_{n_qL}/m_{\bar\chi\chi L}=1.5$ \cite{Nadkarni:1986cz}.}
the screening masses obtained from $\lla n_q^f L\rla$ are larger than those from $\lla \bar\chi\chi^f L \rla_c$ and we generally find  $m^s$ is larger than $m^l$. We observe a drop of all screening masses close to $T_c$, as all correlation functions become large in the cross over region.  For large temperatures we have $m^{l,s}/T\ll 1$ and we therefore expect $m^l=m^s$ to hold, which is not yet the case at $T/T_c=2$.  Moreover, the deviation of the data from the lattice with smaller aspect ratio points to volume effects which are still present in our calculations of $m^{l,s}$.   
%
%
The Debye masses extracted from the $N_\tau=6$ lattices are roughly of the same size as the screening masses \cite{kostya}. 
In order to avoid contamination from excited states, we plan to study $p=0$ projections of the correlators in the future.

The total induced quark number of flavor $f$ around a static anti-quark can be computed by 
\beq
\lla N^f_q\rla_{\overline Q}=\frac{\lla L^\ast \sum_{\bf x} n_q^f({\bf x})  \rla}{\lla \re L\rla},
\eeq
where we show the results of our calculation in fig.~\ref{fig:screening_masses}(right). As was already observed in 2-flavor QCD \cite{Doring:2007uh}, $\lla N^f_q\rla_{\overline Q}$ vanishes rapidly for $T>T_c$, where screening is predominantly provided by thermal gluons. For $T\to 0$ the induced quark number approaches one, where this limit is approached faster for light quarks than the strange flavor, since the heavier sea quarks are more difficult to induce.



%
\begin{figure}
\scalebox{0.6}{\includegraphics{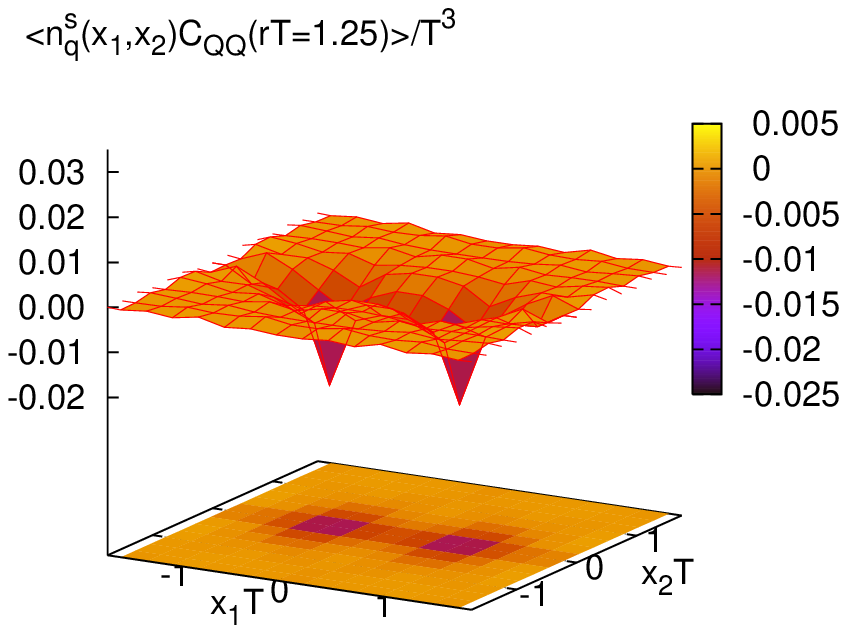}}
\scalebox{0.6}{\includegraphics{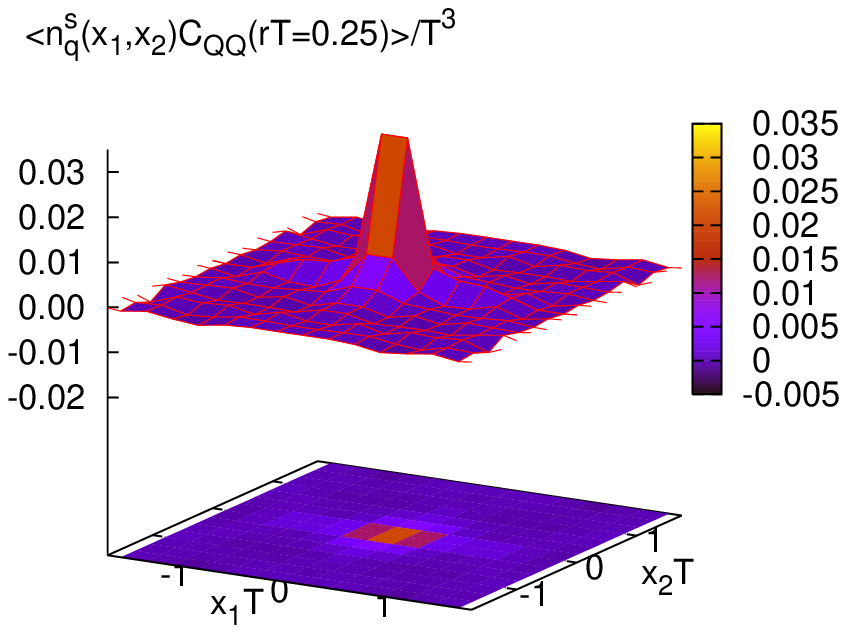}}
\caption{\label{fig:QQ}$\lla n_q^s\,C_{QQ}\rla$ for $T/T_c=0.905, N_\tau=4$ at separations $rT=1.25$ (left) and $rT=0.25$ (right) of the Polyakov loops.
}
\end{figure}
\begin{figure}
\scalebox{0.6}{\includegraphics{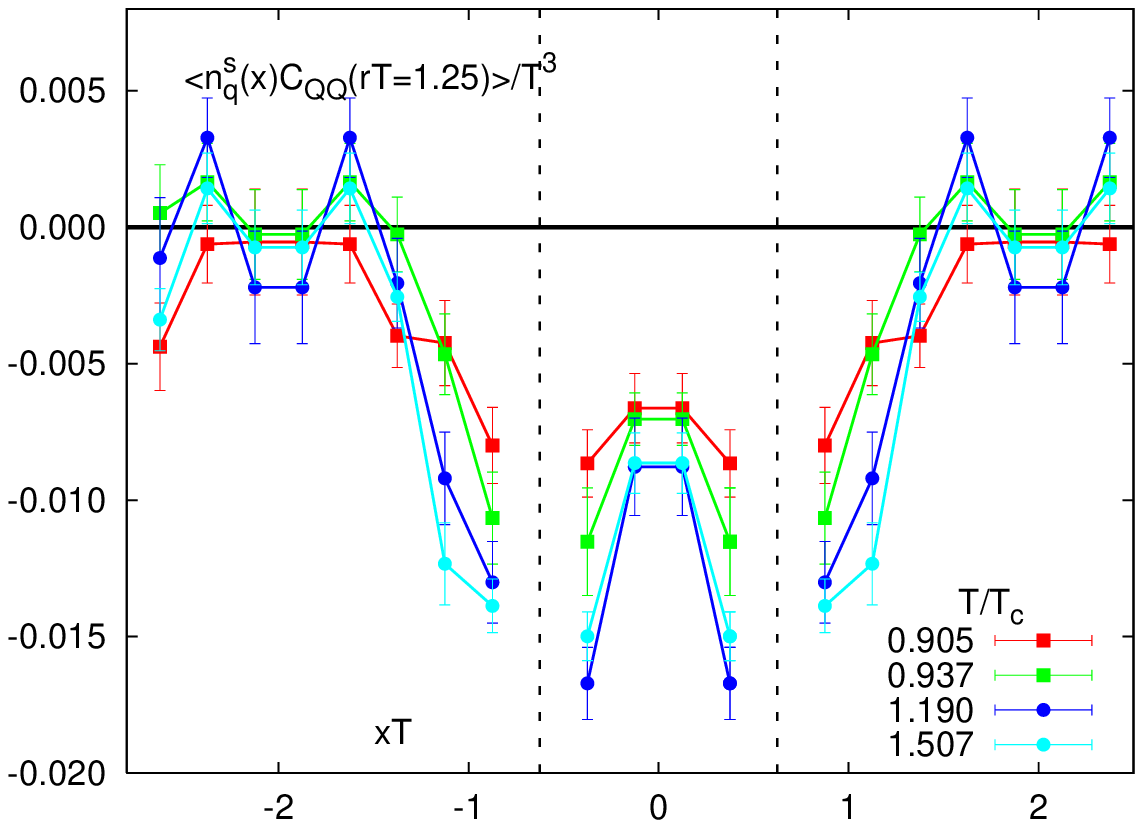}}
\scalebox{0.6}{\includegraphics{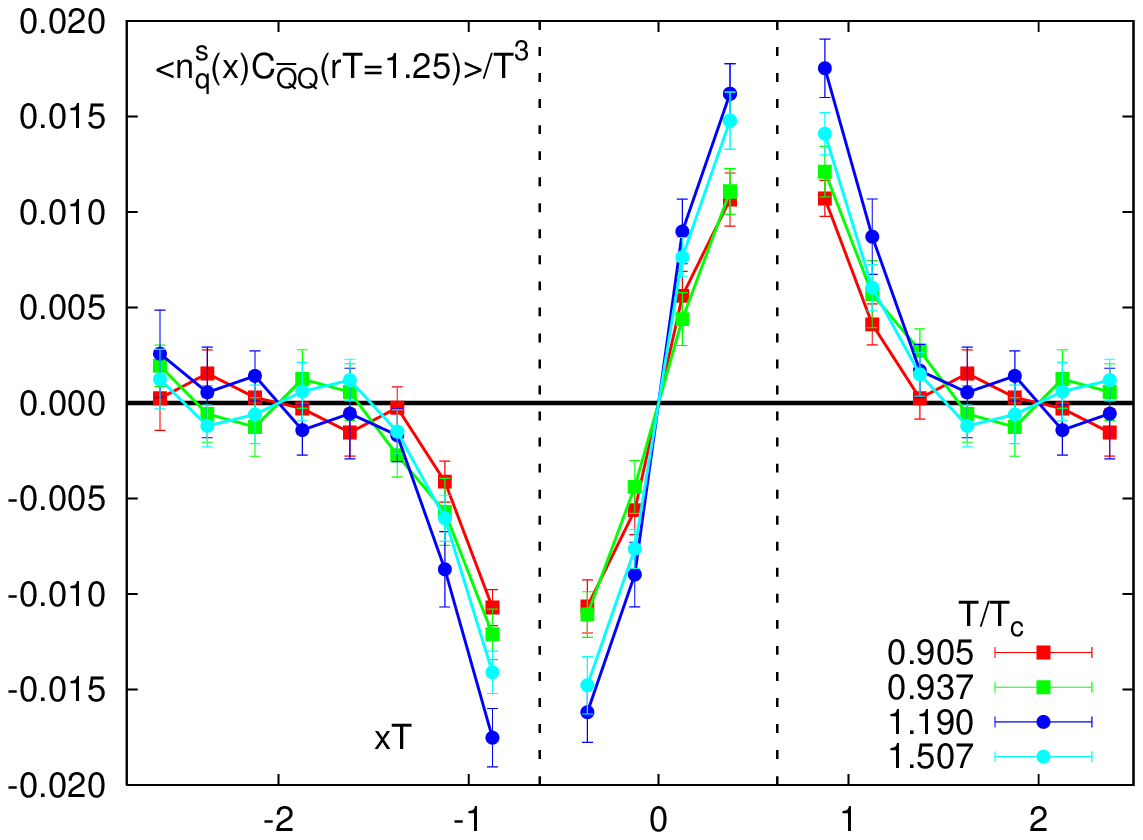}}
\scalebox{0.6}{\includegraphics{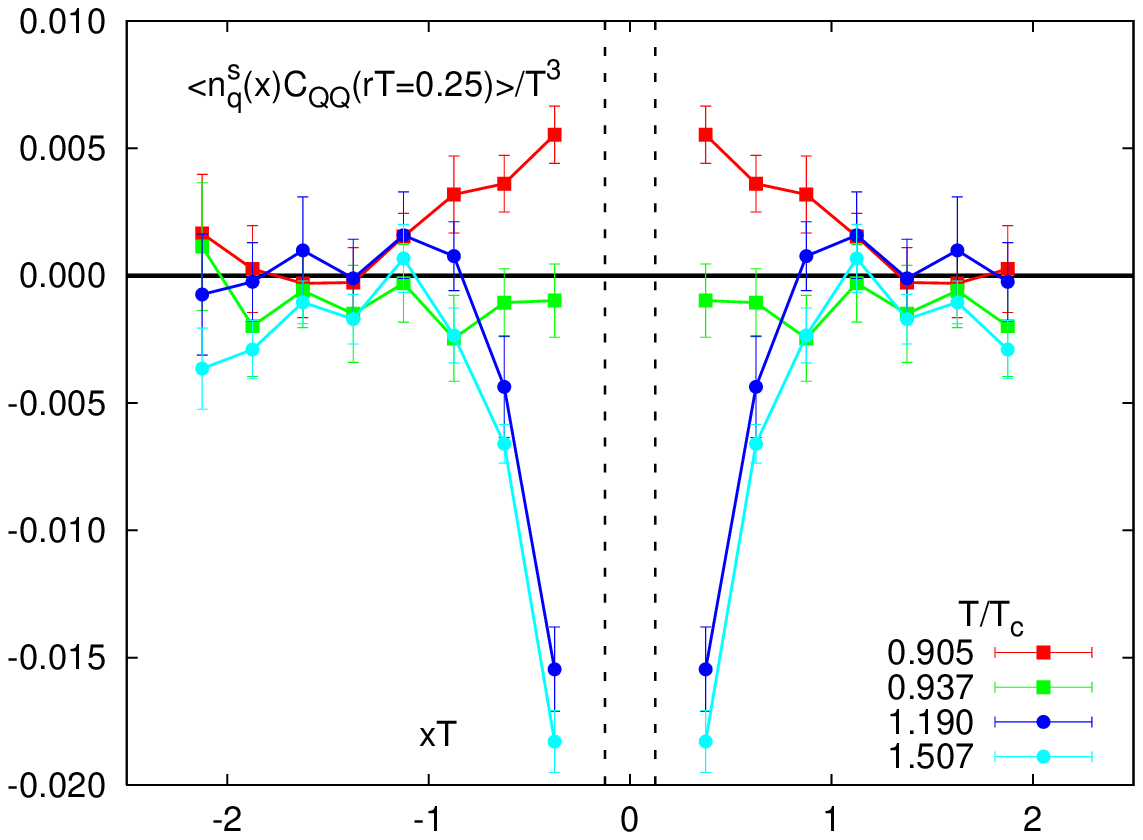}}
\scalebox{0.6}{\includegraphics{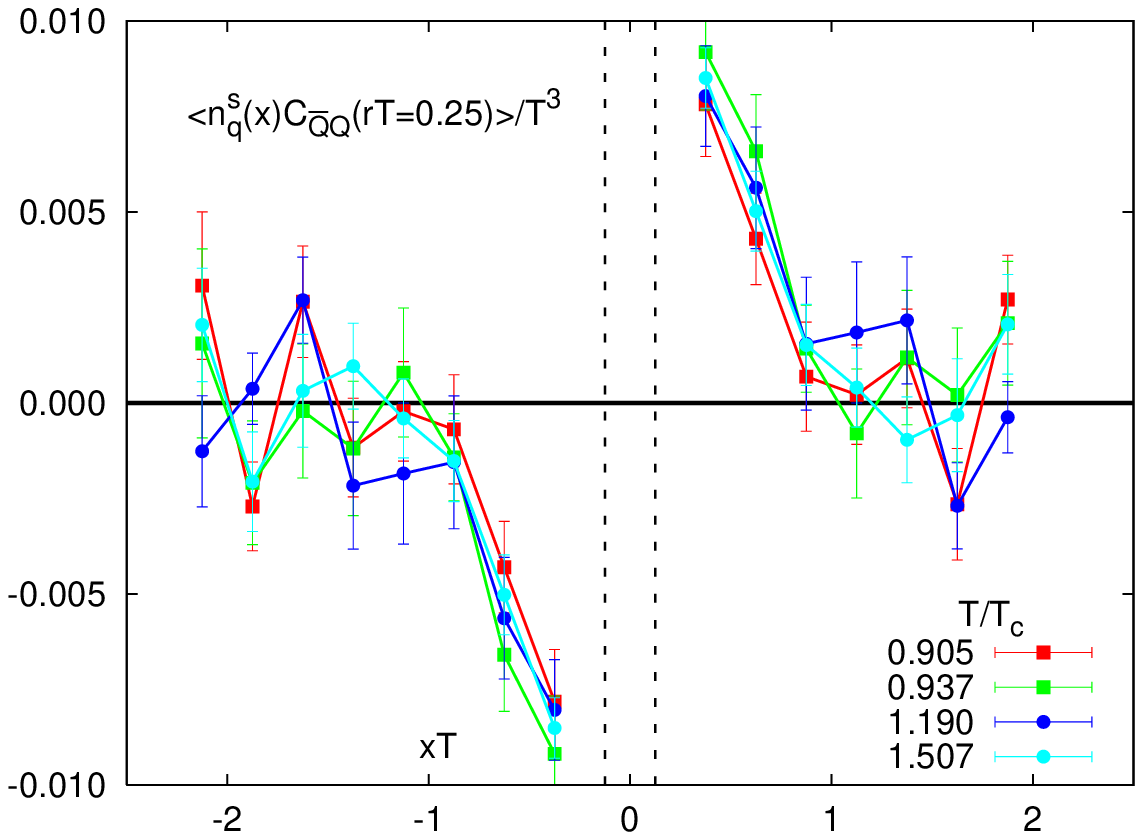}}
\caption{\label{fig:QQ2}Cut along the connecting axis of the static quarks, $\lla n_q^s\,C_{QQ}\rla$ (left) and $\lla n_q^s\,C_{Q\bar Q}\rla$ (right) for several temperatures. The dotted lines give the position of the Polyakov loops.
}
\end{figure}
%

\section{Results for $\lla n_q\,C_{QQ}\rla$ and $\lla n_q\,C_{Q\bar Q}\rla$}

We now turn to the computation of three point correlation functions  $\lla n_q\,C_{QQ}\rla$ and $\lla n_q\,C_{Q\bar Q}\rla$. 
In fig.~\ref{fig:QQ}  we show $\lla n_q^s({\bf x}_1,{\bf x}_2)\,C_{QQ}(r)\rla$ for $T/T_c=0.905$, calculated on lattices with temporal extent $N_\tau=4$ for separations $rT=1.25$ (left) and $rT=0.25$ (right) of the Polyakov loops, 
which are placed at ${\bf x}_1=\pm rT/2,{\bf x}_2=0$.
We see that for large separations $r$ the correlator assumes negative values centered around the individual positions of the static quarks. This signals an excess of dynamical anti-quarks and the screening mechanism is {\em mesonic}, i.~e.~$QQ\to Q\bar q+Q\bar q$, where $Q$ stands for a static quark and $q$ for a dynamical quark. For small separations we observe positive values of the correlator. Thus the screening mechanism has changed to {\em baryonic}, i.~e.~$QQ\to QQq$.
This behavior has been observed in the induced quark number of the diquark system in 2-flavor QCD as well \cite{Doring:2007uh}.

In fig.~\ref{fig:QQ2} we show the values of $\lla n_q^s\,C_{QQ}\rla$ (left) and $\lla n_q^s\,C_{Q\bar Q}\rla$ (right) along the line connecting the static quarks for several temperatures and two separations, where ${\bf x}_1=x,{\bf x}_2=0$. The dotted lines mark the position of the Polyakov loops, where the anti-quark source in the $Q\bar Q$-system is located at larger values of $x$. We see mesonic screening in the diquark system for large separations $r$ at all $T$ calculated here, whereas at the smallest distance only the lowest temperature turns to baryonic screening. In the $Q\bar Q$-system the screening mechanism is mesonic for all $T$ and $rT$, i.~e.~$Q\bar Q\to Q\bar q+\bar Qq$.

%



\section{Conclusion and Outlook}

We have studied correlators of the local chiral condensate and the local quark number density with Polyakov loops in 2+1 flavor QCD with almost realistic quark masses at finite temperature on $16^3\times 4,6$ lattices.
We found that the correlator of the Polyakov loop with the quark number density and  the connected correlator  with the chiral condensate are well described by an exponential fit-Ansatz.
We calculated the screening masses from these correlators and the induced quark number in the static quark system.
Moreover, we investigated the quark number density in static diquark and mesonic systems, where we found the screening mechanism in the diquark system to change from mesonic to baryonic at small separations for small temperatures.

In the future we want to study the $p=0$ projection of the correlators with the Polyakov loop to avoid contamination from excited states in the screening masses. Furthermore we like to examine lattices with a larger physical volume in order to get better control over  finite volume effects.
This work forms the basis for future computations of spatial correlators of the quark number density $\lla n_q(0)n_q(r)\rla$, which might shed light on whether the QGP has a liquid phase \cite{Thoma:2005aw}.


\section{Acknowledgement}

This manuscript has been authored under contract number DE-AC02-98CH10886 with the U.~S.~Department of Energy.


\begin{thebibliography}{99}

\bibitem{Feilmair:1989hi}
  W.~Feilmair, M.~Faber and H.~Markum,
  Phys.\ Lett.\  B {\bf 221} (1989) 363.

\bibitem{Feilmair:1988js}
  W.~Feilmair, M.~Faber and H.~Markum,
  Phys.\ Rev.\  D {\bf 39}, 1409 (1989).

\bibitem{Sakuler:1992qx}
  W.~Sakuler {\it et al.},
  Phys.\ Lett.\  B {\bf 276}, 155 (1992).

\bibitem{Buerger:1994zc}
  W.~Buerger, M.~Faber, H.~Markum, M.~Muller and M.~Schaler,
  Nucl.\ Phys.\ Proc.\ Suppl.\  {\bf 34}, 269 (1994).

\bibitem{Thoma:2005aw}
  M.~H.~Thoma,
  Nucl.\ Phys.\  A {\bf 774}, 307 (2006)


\bibitem{EoS}
Jan van der Heide, ``Thermodynamics of 2+1 flavour QCD``, this Proceedings

\bibitem{Cheng:2006qk}
  M.~Cheng {\it et al.},
  Phys.\ Rev.\  D {\bf 74}, 054507 (2006)


\bibitem{Nadkarni:1986cz}
  S.~Nadkarni,
  Phys.\ Rev.\  D {\bf 33} (1986) 3738.

\bibitem{kostya}
K.~Petrov, ``Entropy and Internal energy of Heavy quark-anti-quark pair``,  this Proceedings




\bibitem{Doring:2007uh}
  M.~Doring, K.~Hubner, O.~Kaczmarek and F.~Karsch,
  Phys.\ Rev.\  D {\bf 75}, 054504 (2007)




\end{thebibliography}
\end{document}